\documentclass[twoside,fleqn]{article}
\usepackage{latexsym,espcrc2}

\title{   High-precision computation of two-loop Feynman diagrams
          with Wilson fermions}

\author{
  Stefano Capitani\address{DESY, Notkestra\ss e 85,  D-22607 Hamburg, 
  Germany},
  Sergio Caracciolo\address{Scuola Normale Superiore and INFN,
      Sezione di Pisa,
      I-56100 Pisa, Italy},
   Andrea Pelissetto\address{
      Dipartimento di Fisica and INFN,
      Universit\`a degli Studi di Pisa,
      I-56100 Pisa, Italy}
  and
 Giancarlo Rossi\address{Dipartimento di Fisica, Universit\`a  di 
 Roma ``Tor Vergata'' and INFN, Sezione di Roma II, Via della Ricerca 
 Scientifica 1, I-00133 Roma, Italy}
}
\begin{document}
\begin{abstract}
We apply the coordinate-space method by L\"{u}scher and Weisz  
to the computation 
of two-loop diagrams in full QCD with Wilson fermions on the lattice. 
The essential ingredient
is the high-precision determination of mixed fermionic-bosonic propagators.
%%%%%%%%%%%%%%%%%%%%%%%%%%%%%%%%%%%%%%%%%%%%%%%%%%%%%%%%%%%%%%%%%%%%%%%%%%%%%
%						
% As an example we present some preliminary results on the calculation of the
% mixing coefficients of $\overline{\psi} \sigma \cdot F \psi$ with the weak
% hamiltonian which is relevant for the evaluation of the $\Delta I = 1/2$
% weak matrix elements.
%
%%%%%%%%%%%%%%%%%%%%%%%%%%%%%%%%%%%%%%%%%%%%%%%%%%%%%%%%%%%%%%%%%%%%%%%%%%%%%
\end{abstract}

\maketitle

\newcommand{\be}{\begin{equation}}
\newcommand{\ee}{\end{equation}}
\newcommand{\<}{\langle}
\renewcommand{\>}{\rangle}

\newcommand{\reff}[1]{(\ref{#1})}

At the Lattice conference of last year we presented~\cite{lat96} an algebraic 
algorithm
that allows to express every one-loop lattice integral with gluon or 
Wilson-fermion 
propagator in terms of a small number of basic constants which can be computed 
with 
arbitrary high precision~\cite{BCP}.
This was a generalization of what we did previously with purely bosonic 
integrals~\cite{CMP} and it was also an essential step in order to apply 
the coordinate-space method  by L\"uscher and Weisz~\cite{LW} to higher-loop 
integrals with fermions.

Let us consider, in order to fix the ideas,  an example of a two-loop integral 
at zero external momentum, like
\be
I = \int \,{d^4l\over (2\pi)^4}\,{d^4r\over (2\pi)^4}\, 
            {1\over D_F(l) D_F(r) D_F(l+r)}
\label{defI}
\ee
where
\be
D_F(l) = \sum_{i=1}^4 \sin ^2 l_i  + {r_W\over 4} (\hat{l}^2)^2
\ee
defines 
the usual Wilson propagator with momentum $\vec{l}$ at zero fermion mass 
with Wilson parameter $r_W$. In all our calculation we have set 
$r_W = 1$ but the method can be applied for any value of $r_W$.
A possible
strategy to evaluate such an integral amounts to replace each integration
with a discrete sum over $L$ points and afterwards to extrapolate to infinite 
$L$.
A possibility is rewriting \reff{defI} as 
\be
I = {1\over L^8} \sum_{l,r,l+r\not= 0} {1\over D_F(l) D_F(r) D_F(l+r)}
\label{defsum}
\ee
where each component $l_i$ and $r_i$ runs over the set 
$2\pi (n + 1/2)/L$, $n=0,\ldots,L-1$. From the sum we exclude 
the points such that $l+r=0$ $\bmod{2\pi}$ where the third
propagator  diverges. Another possibility is to use \reff{defsum}
but with 
$l_i$ and $r_i$ running over the set 
$2\pi n/L$, $n=0,\ldots,L-1$. In this case however there are more problems with
the zero modes and one should exclude from the sum the terms 
with $l=r=l+r=0$ mod $2\pi$. For this reason we have decided to use 
the first method.
For our previous example we get for increasing values of $L$
\begin{eqnarray*}
L=10 &\quad & 0.000 799 652  \\
L=18 &\quad & 0.000 848 862  \\
L=20 &\quad & 0.000 853 822  \\
L=26 &\quad & 0.000 863 064  
\end{eqnarray*}
Then, using an extrapolation of the form 
\be
a_0 + {a_1 \log L + a_2\over L^2} +
      {a_3 \log L + a_4\over L^4} 
\ee
and data with $6\le L \le 26$, we obtain the estimate
$$
I \approx 0.000 879 776
$$
which has to be compared 
with what we obtained by using the coordinate-space method
$$
I \approx 0.000 879 777 918 1 (12)
$$
Let
\be
G(p,q,\vec{x}) = \int \, dk {e^{i \vec{k}\cdot\vec{x}}\over D_F^p(k) D_B^q(k)}
\ee
where $D_B(k) = \hat{k}^2$ is the usual bosonic propagator on the lattice.
In the coordinate-space approach we are interested in the evaluation of
the lattice sums
\begin{eqnarray}
\lefteqn{I(p_1,q_1,\vec{a},p_2,q_2,\vec{b},p_3,q_3,\vec{c})=}\label{base}
\\
& & \hspace{-0.5cm}
 \sum_{\vec{x}} G(p_1,q_1,\vec{x}+ \vec{a})  G(p_2,q_2,\vec{x}+ \vec{b}) 
                      G(p_3,q_3,\vec{x}+ \vec{c})  \nonumber 
\end{eqnarray}
In this notation our previous example corresponds to
\be
I = I(1,0,\vec{0},1,0,\vec{0},1,0,\vec{0})
\ee
In the evaluations of these sums we make use of the following advantages:
\begin{itemize}
\item only four infinite lattice sums  must be computed;
\item the $G(p,q,\vec{x})$'s can be determined with the desired precision, 
for a 
sufficiently large domain of values of $\vec{x}$, by using our algebraic 
algorithm~\cite{BCP};
\item the asymptotic expansion for large values of $|\vec{x}|$ of the
$G(p,q,\vec{x})$'s is easily computed. For example
\begin{eqnarray}
\lefteqn{G(1,0,\vec{x}) = {1\over \pi^2} \left[ {1\over 4 x_2} - {1\over x_2^2} + 
{2 x_4\over x_2^4} - {10\over x_2^3} + \right.} \nonumber \\
& &\left.  +{52\over x_2^5} + {160 x_4^2\over x_2^7} - {192 x_6\over x_2^6} + 
\ldots \right]
\end{eqnarray}
where $x_n = \sum_\mu x_\mu^n$.
\end{itemize}
Let us now consider how to compute sums of the type 
\be 
   \Sigma = \sum_{\Lambda} f(x)
\ee
on the lattice $\Lambda$.
Of course we will not be able to sum over all the lattice. 
If $|x|_1 = \sum_{\mu} |x_{\mu}|$, we will perform a sum over 
a domain of the type $D_p = \{x\in \Lambda: |x|_1 \le p\}$. The problem is 
to give an estimate of the error. If $f(x)$ decreases for large $|x|$
as $1/|x|^{2 k}$ we expect the sum restricted to $D_p$ to behave as 
\be 
   \Sigma(p) =\sum_{D_p} f(x) = \Sigma + {A\over p^{2k - 4}} + \ldots
\ee
Thus we will estimate 
\be
   |\Sigma - \Sigma(p)| = {p\over 2k-4} |\Sigma(p) - \Sigma(p-1) |
\ee
Our error formula seems to work correctly. We will use this formula 
to estimate the error on the integration sums. We can also define 
an improved estimate for $\Sigma$ by 
\be
\Sigma \approx \Sigma(p) + {p\over 2k - 4} (\Sigma(p) - \Sigma(p-1))
\ee
Notice that now our error estimate is {\em very} conservative.
The larger is $k$ the best is the estimate. For this reason, if we 
know the asymptotic behaviour $Af(x)$ of the function $f(x)$ for large $x$ 
it is convenient to write
\be
\Sigma \approx \sum_{D_{p}} \left[ f(x) - Af(x) \right] + 
              \sum_{\Lambda} Af(x)
\ee
because the difference is decreasing faster at infinity and the sum 
of $Af(x)$ can be computed directly on the infinite lattice by using 
harmonic polynomials and $\zeta$-functions as explained in 
\cite{LW}.
Coming back to our preferred example, by subtracting an 
increasing number of terms of the asymptotic expansion, we get the estimates
\begin{eqnarray*}
I  &=&  0.0008798104043 \pm
                  0.0000008730034  \\
                &=&  0.0008797776858 \pm
                  0.0000000029778  \\
                &=&  0.0008797779227 \pm
                   0.0000000000410  \\
                &=&  0.0008797779181  \pm
                   0.0000000000012  
\end{eqnarray*}
In our work we have always used the asymptotic expansions to increase the 
precision of the estimates. In each case we have subtracted the asymptotic
behaviour to order $1/|x|^{10}$: therefore the function which is summed 
over a finite lattice decays at least as $1/|x|^{12}$.

A number of checks have been performed on the table of numerical 
integrals that we have collected (at the moment we  have a number of entries of 
order $10^{4}$). In particular
\begin{itemize}
\item In the case in which we restrict ourselves to purely bosonic integrals we 
compare perfectly with the numbers given in~\cite{LW}.
\item Because of translation invariance, for every $\vec{v}$ on the 
lattice
\begin{eqnarray}
\lefteqn{
I(p_1,q_1,\vec{a},p_2,q_2,\vec{b},p_3,q_3,\vec{c}) =} \\
& &I(p_1,q_1,\vec{a}+\vec{v},p_2,q_2,\vec{b}+\vec{v},p_3,q_3,\vec{c}+\vec{v})
\nonumber
\end{eqnarray}
\item From the definition of the bosonic propagator one easily gets
\be{1\over D_{B}^{p}(k)} = {\sum_{\mu} \left( 2 - e^{i k_{\mu}} - e^{- i 
k_{\mu}} \right) \over D_{B}^{p+1}(k)}
\ee
which can be used to derive relations among different $I$'s. 
Similarly, other relations can be obtained from the definition of the 
fermionic propagator.
\item New identities are obtained by integration by parts, that is by
using
\be
\int\,dk_{\mu} {\partial \over \partial k_{\mu}} F\left( k_{\mu} 
\right) = 0
\ee
\end{itemize}
All the checks we have performed are satisfied with a precision of at 
least $10^{-10}$

We have prepared a completely automatic procedure which evaluates 
Feynman diagrams at two loops. It goes through the following steps
\begin{itemize}
\item[1.] Each diagram is reduced as a sum of the 
integrals $I$'s, previously defined in~\reff{base}. 
A mass is added at all propagators in order to 
regularize the infrared divergences.
\item[2.] All the possible symmetries (cubic: translations, 
permutations of the axes, inversions of the axes; permutations of 
the three propagators) are used to reduce the number of terms.
\item[3.] Through subtractions all the terms are written as 
convergent sums plus product of 1-loop integrals.
\item[4.] All 1-loop integrals are expressed as in \cite{lat96,BCP}.
\item[5.] The convergent sums are replaced by their numerical
estimate obtained from a precompiled table of lattice sums in the
domain $D_{21}$.
\end{itemize}
This procedure is now being used to compute the mixing coefficients
of the four fermion operators of the lattice weak hamiltonian with
the dimension 5 operators. We have chosen this computation because
it starts at the two-loop level and because these coefficients
have already been studied by using the momentum-space
approach~\cite{Curci}. The computation is highly non trivial.
The typical input, for each diagram, contains order $10^4$ terms
and produces a final result of order $10^{-4}$.
Unfortunately at the moment  our results are still preliminary and we
are performing all possible checks on our evaluations.

\end{document}